\documentclass[aps,prb,twocolumn,showpacs,amsmath,amssymb]{revtex4}
\usepackage{graphicx}% Include figure files
\usepackage{dcolumn}% Align table columns on decimal point
\usepackage{bm}% bold math

\begin{document}
\title{Wannier functions and exchange integrals: The example of LiCu$_{2}$O$_{2}$}
\author{V.V. Mazurenko$^{1,2}$, S.L. Skornyakov$^{1}$, A.V. Kozhevnikov$^{3}$, F. Mila$^{2}$ and V.I. Anisimov$^{1,3}$}
\affiliation{$^{1}$Theoretical Physics and Applied Mathematics Department, Urals State Technical University, Mira Street 19,  620002
Ekaterinburg, Russia \\
$^{2}$Institute of Theoretical Physics, Swiss Federal Institute of Technology (EPFL), 
CH-1015 Lausanne, Switzerland \\
$^{3}$Institute of Metal Physics, Russian Academy of Sciences, 620219 Ekaterinburg GSP-170, Russia}
\date{\today}

\begin{abstract}
Starting from a single band Hubbard model in the Wannier function basis, we revisit the problem
of the ligand contribution to exchange and derive explicit formulae for the exchange integrals
in metal oxide compounds in terms of atomic parameters that can be calculated with
constrained LDA and LDA+U.
The analysis is applied to the investigation of the isotropic exchange interactions of LiCu$_{2}$O$_{2}$,
a compound where the Cu-O-Cu angle of the dominant exchange path is close to 90$^{\circ}$.
Our results show that the magnetic moments are localized in Wannier orbitals which have strong 
contribution from oxygen atomic orbitals, leading to exchange integrals that considerably
differ from the estimates based on kinetic exchange only. Using
LSDA+U approach, we also perform a direct {\it ab-initio} determination
of the exchange integrals LiCu$_{2}$O$_{2}$. The results agree well with those obtained
from the Wannier function approach, a clear indication that this modelization captures
the essential physics of exchange. A comparison with experimental results is also
included, with the conclusion that a very precise determination of the Wannier
function is crucial to reach quantitative estimates.
\end{abstract}

\maketitle

\section{Introduction}
In order to explain the homeopolar chemical bond in the hydrogen molecule, 
Heitler and London \cite{Heitler} have derived the direct 
and exchange Coulomb integrals which can be written in the following form:
\begin{eqnarray}
U = \int  \frac{\phi^{*}_{i} (x) \phi_{i} (x) \phi^{*}_{i} (x') \phi_{i} (x')}{|x-x'|} dx dx' 
\end{eqnarray}
and 
\begin{eqnarray}
J^{Coulomb}_{ij} = \int  \frac{\phi^{*}_{i} (x) \phi_{j} (x) \phi^{*}_{j} (x') \phi_{i} (x')}{|x-x'|} dx dx', 
\end{eqnarray}
where {\it i} and {\it j} are site indexes, and $\phi_{i} (x)$ is a wave function centered at the lattice site {\it i}. 
Taking into account the exchange Coulomb integral allows one to obtain bonding ($E=U-J^{Coulomb}_{ij}$) 
as well as antibonding  
($E'=U+J^{Coulomb}_{ij}$) states of H$_{2}$.
Heitler and London have shown that the properties of the 
hydrogen molecule can be described correctly using these combinations of wave functions.

Then, in 1928, Heisenberg \cite{Heisenberg} has used the exchange Coulomb integral $J^{Coulomb}_{ij}$ 
to explain ferromagnetism. Heisenberg has supposed that this exchange Coulomb 
integral corresponds to the exchange 
coupling in a spin model defined by the Hamiltonian:
\begin{equation}
H=\sum_{ij} J_{ij} \vec S_{i} \cdot \vec S_{j},
\end{equation} 
and that it is the main source of ferromagnetism in
3d metal compounds. However, if $\phi_{i} (x)$ in Eq.(1) is the atomic wave function centered 
on the $ith$ atom, then the overlap between wave functions 
on neighboring atoms is negligibly small. Therefore the exchange interaction derived 
through Eq.(2) can be neglected.    

In 1959, P.W. Anderson \cite{Anderson} has suggested a new type of exchange interaction 
process based on hopping (kinetic exchange interaction). 
In the context of the Hubbard model, \cite{Hubbard}
\begin{equation}
H=\sum_{ij \sigma} t_{ij} a^{+}_{i \sigma} a_{j \sigma} + 
\frac{U}{2} \sum_{i \sigma} n_{i \sigma} n_{i -\sigma},
\end{equation}
this exchange interaction can be expressed as 
$J^{kin}_{ij}=\frac{2 t_{ij}^{2}}{U}$. This estimation of exchange interaction through 
hopping integrals is convenient and now widely used in the
literature. However, if $J^{kin}_{ij}$ is small enough, other sources of exchange coupling become
important. 
Indeed, the proper way to discuss exchange is to consider the basis of Wannier functions $W_{i}$ 
(where $i$ is the composite index of band and site) proposed by Wannier \cite{wannier} in 1937 
and defined as the Fourier transforms
of a certain linear combination of Bloch functions $\psi_{nk}$ ($n$ is the band index and $k$ is 
the wave-vector in reciprocal space). In contrast to atomic wave functions $\phi_{i} (x)$, 
which are localized 
on one atom, the 
Wannier functions $W_{i}(x)$ are more extended in space and can be expressed through linear 
combination atomic wave functions, 
$W_{i}(x)=\sum_{j T} \alpha^{i}_{j} \phi_{j}(x-T)$ (where $\alpha^{i}_{j}$ is the contribution of the {\it j}th 
atomic orbital to Wannier
function $W_{i}(x)$ and $T$ is a translation vector). The 
Wannier functions are the most localized ones within the subspace of low-energy excitations, 
which facilitates a
direct physical interpretation consistent with interacting localized spins. Therefore, to use 
Wannier states instead of atomic sets is physically motivated. In particular, as we shall see,
$J^{Coulomb}_{ij}$ defined in the Wannier basis plays a crucial role in the 
description of exchange interactions between magnetic moments in the case of nearly 90$^{\circ}$ 
metal-oxygen-metal bonds. 

These questions have already been discussed in several context in the literature.
For instance, the authors of Ref. \onlinecite{Kahn} have discussed two alternative ways, natural
or orthogonalized magnetic orbitals, to describe the exchange interactions. They have concluded 
that the orthogonalized magnetic orbital approach clearly
leads to simpler calculations and thus may be more appropriate for quantitative computations. 

The competition between kinetic ($J^{kin}_{ij}$) and potential ($J^{Coulomb}_{ij}$)
contributions to the total exchange interaction has been considered before in many works. 
In Ref. \onlinecite{Graaf}, the authors have performed model calculations of a
Cu$_{2}$O$_{6}$Li$_{4}$ cluster for different
Cu-O-Cu angles. Their computational experiments have shown that the 
nearest-neighbor interaction reaches a maximum around 97$^{\circ}$ and remains
ferromagnetic up to angles as large as 104$^{\circ}$. They have also concluded that the 
simple superexchange relation cannot be
applied to Li$_{2}$CuO$_{2}$. In Ref. \onlinecite{Pickett}, it was shown that unusual 
insulating ferromagnetism in La$_{4}$Ba$_{2}$Cu$_{2}$O$_{10}$ can be 
explained by intersite ferromagnetic "direct exchange" (in our notation $J^{Coulomb}_{ij}$). 
The authors have concluded that the latter
process occurs mainly at La and O sites and overwhelms the AF superexchange J$^{kin}_{ij}$. 
The value of J$^{Coulomb}_{ij}$ was
calculated through a direct integration over the wave functions.
  
In the present paper, we revisit this issue in the context of constrained LDA and LSDA+U
{\it ab-initio} approaches. The problem we want to address can be formulated as follows. 
In favourable situations, the exchange integrals calculated using LSDA+U agree quite
well with the standard superexchange expression $J^{kin}_{ij}=\frac{2 t_{ij}^{2}}{U}$
if the parameters $t_{ij}$ and $U$ are themselves determined from constrained LDA and 
LSDA+U.
Such an interpretation of the LSDA+U results is 
a well accepted criterion to test the reliability of the result. But 
the standard superexchange expression $J^{kin}_{ij}=\frac{2 t_{ij}^{2}}{U}$
often fails quantitatively, and even sometimes qualitatively, to reproduce the LSDA+U results.
The main goal of this paper is to provide for such cases a generalization of the 
superexchange expression which is entirely expressed in terms of parameters 
that can be determined by constrained LDA and LSAD+U, and which can be used as an interpretation and
an independent check of  the LSDA+U results.

To this end, we start from the standard Hamiltonian in Wannier function basis for nearly 
90$^{\circ}$ metal-oxygen-metal bond. We show how the different Coulomb interaction 
terms of this Hamiltonian are related to parameters defined in the atomic basis set,
with emphasis on the intraatomic exchange interaction of oxygen, J$_{p}^{H}$,
to which $J^{Coulomb}_{ij}$ is proportional 
when neighboring Wannier orbitals overlap on the
oxygen atoms. We then present a simple expression for the 
exchange interactions between magnetic moments in the system. 
The parameters that enter this expression  can themselves 
be estimated through LDA and LSDA+U calculations. 
This formalism is applied to the investigation of the magnetic properties
of LiCu$_{2}$O$_{2}$, and the results are compared to those 
of first-principle LSDA+U approach.
   
The paper is organized as follows. In Section II, we discuss the Hubbard model in Wannier function basis. 
In Section III A, we shortly describe the crystal structure of LiCu$_{2}$O$_{2}$ and present 
the results of LDA calculation. In Section III B, we present the results of LSDA+U calculations and discuss
the exchange interactions obtained between different pairs of magnetic moments in LiCu$_{2}$O$_{2}$.
In Section IV, we briefly summarize our results.   

\section{Hubbard model in Wannier function basis}
The general Hamiltonian in Wannier function basis $W_{i}(x)$ can be written in the following form: \cite{Hubbard}
\begin{eqnarray}
H = \sum_{i, j, \sigma} t_{ij} a^{+}_{i \sigma} a_{j \sigma} + \frac{1}{2} \sum_{i j k l, \sigma \sigma'}
 (ij|U|kl) a^{+}_{i \sigma} a^{+}_{j \sigma'} a_{l \sigma'} a_{k \sigma},
\end{eqnarray}
where 
\begin{eqnarray}
t_{ij} = \int W^{*}_{i} (x) \nabla^{2} W_{j} (x) dx \nonumber 
\end{eqnarray}
and 
\begin{eqnarray}
(ij|U|kl) = \int \frac{W^{*}_{i} (x) W_{k} (x) W^{*}_{j} (x') W_{l} (x')}{|x-x'|} dx dx'. \nonumber 
\end{eqnarray}

We analyze the complex Hamiltonian of Eq.(5) in the context of a
3-site model (see Fig.1) with nearly 90$^{\circ}$ bonds and define two Wannier orbitals  
$W_{1} = \alpha \phi_{1} + \beta \phi_{p_1}$ and 
$W_{2} = \alpha \phi_{2} + \beta \phi_{p_2}$,
which are constructed from the atomic wave functions: $\phi_{1}$, $\phi_{2}$, $\phi_{p_1}$ and $\phi_{p_2}$ (see Fig.1).
\begin{figure}[!h]
\includegraphics[width=0.33\textwidth]{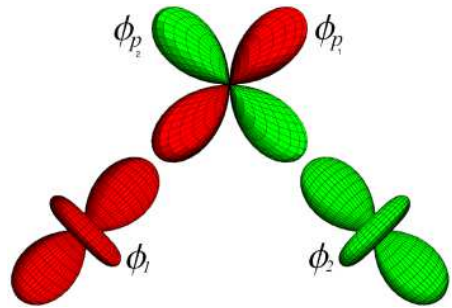}
\caption{ (Color online) Schematic representation of a nearly 90$^{\circ}$ bond between 3d atoms through oxygen.}
\label{bandslda}
\end{figure}
The oxygen orbitals $\phi_{p_1}$ and $\phi_{p_2}$ can be expressed through the angle of metal-oxygen-metal 
bond $\theta$ in the following form:
\begin{eqnarray}
\phi_{p_1} = \cos (\frac{\theta - 90^{\circ}}{2}) \phi_{p_{y}} + \sin (\frac{\theta - 90^{\circ}}{2}) \phi_{p_{x}}  
\end{eqnarray}
and 
\begin{eqnarray}
\phi_{p_2} = \cos (\frac{\theta - 90^{\circ}}{2}) \phi_{p_{x}} + \sin (\frac{\theta - 90^{\circ}}{2}) \phi_{p_{y}}.  
\end{eqnarray}

Let us first analyze the hopping term in Eq.(5).
It is easy to show that 
\begin{eqnarray}
t_{12} = \int W^{*}_{1} (x) \nabla^{2} W_{2} (x) dx = \beta^{2} \int \phi^{*}_{p_1} (x) \nabla^{2} \phi_{p_2} (x) dx.  
\end{eqnarray}
If $\phi_{p_1}$ and $\phi_{p_2}$ are eigenfunctions of the Hamiltonian of Eq.(5), then
using the orthogonality condition, one can obtain the following expression for the hopping term:
\begin{eqnarray}
t_{12} \sim \beta^{2} \cos \theta.  
\end{eqnarray}
Clearly, if $\theta=90^{\circ}$, the hopping integral vanishes.

In the analysis of the Coulomb term of Eq.(5) that follows, we consider only 
density-density terms and on-site exchange integrals.
On-site Coulomb interaction, intersite Coulomb interaction and intersite 
exchange interaction in Wannier function basis 
are expressed through the corresponding parameters in the atomic basis set:
\begin{eqnarray}
(ii|U|ii) = \alpha^{4} U_{d} + 2 \alpha^{2} \beta^{2} V_{pd} + \beta^{4} U_{p}, \quad \quad \quad \quad \quad \quad 
\end{eqnarray} 
\begin{eqnarray}
(ij|U|ij) = \alpha^{4} V_{dd} + 2 \alpha^{2} \beta^{2} V_{pd}  +\beta^{4} U_{p}, \quad \quad \quad \quad \quad \quad
\end{eqnarray}
\begin{eqnarray}
(ij|U|ji) = \beta^{4} J^{H}_{p} \quad \quad \quad \quad \quad \quad \quad \quad \quad \quad \quad \quad \quad \quad
\end{eqnarray}
where
\begin{eqnarray}
U_{d} = \int  \frac{\phi^{*}_{1} (x) \phi_{1} (x) \phi^{*}_{1} (x') \phi_{1} (x')}{|x-x'|} dx dx' 
\end{eqnarray}
is the on-site Coulomb interaction of 3d atom,
\begin{eqnarray}
U_{p} = \int  \frac{\phi^{*}_{p_1} (x) \phi_{p_1} (x) \phi^{*}_{p_2} (x') \phi_{p_2} (x')}{|x-x'|} dx dx' 
\end{eqnarray}
is the on-site Coulomb interaction of oxygen atom,
\begin{eqnarray}
V_{pd} = \int  \frac{\phi^{*}_{1} (x) \phi_{1} (x) \phi^{*}_{p_1} (x') \phi_{p_1} (x')}{|x-x'|} dx dx' 
\end{eqnarray}
is the Coulomb interaction between 3d atom and oxygen,
\begin{eqnarray}
V_{dd} = \int  \frac{\phi^{*}_{1} (x) \phi_{1} (x) \phi^{*}_{2} (x') \phi_{2} (x')}{|x-x'|} dx dx' 
\end{eqnarray}
is the Coulomb interaction between 3d atoms,
\begin{eqnarray}
J^{H}_{p} = \int  \frac{\phi^{*}_{p_1} (x) \phi_{p_2} (x) \phi^{*}_{p_2} (x') \phi_{p_1} (x')}{|x-x'|} dx dx' 
\end{eqnarray}
is the intraatomic exchange interaction of oxygen atom.

It was shown in Ref. \onlinecite{Cox} that the following term: 
\begin{eqnarray}
(ii|U|ij) = \alpha^{2} \beta^{2} \int \frac{\phi^{*}_{1} (x) \phi_{1} (x) \phi^{*}_{p_1} (x') \phi_{p_2} (x')}{|x-x'|} dx dx'
\nonumber \\
+ \beta^{4} \int \frac{\phi^{*}_{p_1} (x) \phi_{p_1} (x) \phi^{*}_{p_1} (x') \phi_{p_2} (x')}{|x-x'|} dx dx' \quad \quad
\end{eqnarray}
which is the so-called correlated hybridization, could significantly change the parameters in the
effective single band model for transition
metal oxides. However we do not consider it here and leave that point for further investigation.

Finally one can write the following Hamiltonian in Wannier function basis 
\begin{eqnarray}
H &=& \sum_{i, j, \sigma} t_{ij} a^{+}_{i \sigma} a_{j \sigma}  \nonumber \\ 
&+&\frac{(\alpha^{4} U_{d} + 2 \alpha^{2} \beta^{2} V_{pd} + \beta^{4} U_{p})}{2} 
\sum_{i \sigma} n_{i \sigma} n_{i -\sigma} \nonumber \\ 
&+& \frac{ (\alpha^{4} V_{dd} +  2 \alpha^{2} \beta^{2} V_{pd} + \beta^{4} U_{p})}{2} \sum_{i j \sigma \sigma'} n_{i \sigma} n_{j \sigma'} 
\nonumber \\ 
&-& \frac{\beta^{4} J^{H}_{p}}{2} \sum_{i j \sigma} 
(n_{i \sigma} n_{j \sigma} + 2 S^{x}_{i} S^{x}_{j} + 2 S^{y}_{i} S^{y}_{j}), 
\end{eqnarray}
where $n_{i \sigma} = a^{+}_{i \sigma} a_{i \sigma}$ is the particle number operator, while
$S^{x}_{i} = \frac{1}{2} (a^{+}_{i \uparrow} a_{i \downarrow} + a^{+}_{i \downarrow} a_{i \uparrow})$ and 
$S^{y}_{i} = \frac{1}{2 i} (a^{+}_{i \uparrow} a_{i \downarrow} - a^{+}_{i \downarrow} a_{i \uparrow})$ are 
components of the spin operator.
One can reduce Eq.(19) to the following form:
\begin{eqnarray}
H = \sum_{i, j, \sigma} t_{ij} a^{+}_{i \sigma} a_{j \sigma} + \frac{U_{eff}}{2} 
\sum_{i \sigma} n_{i \sigma} n_{i -\sigma} \nonumber \\ + \frac{V_{eff}}{2} \sum_{i j \sigma \sigma'} n_{i \sigma} n_{j \sigma'} 
- \beta^{4} J^{H}_{p} \sum_{i j} \vec {S_{i}} \cdot \vec {S_{j}},  
\end{eqnarray}
where the effective on-site Coulomb interaction is given by $U_{eff} = \alpha^{4} U_{d} + 2 \alpha^{2} \beta^{2} V_{pd} + \beta^{4} U_{p}$ and the effective
intersite Coulomb interaction by
$V_{eff} = \alpha^{4} V_{dd} +  2 \alpha^{2} \beta^{2} V_{pd} + \beta^{4} U_{p} - \frac{\beta^{4} J^{H}_{p}}{2}$.
It is easy to show that the Heisenberg model which corresponds to this electronic Hamiltonian has the following form:
\begin{eqnarray}
H = \sum_{i j} (J^{kin}_{ij} + J^{Coulomb}_{ij}) \vec {S_{i}} \cdot \vec {S_{j}} \nonumber \\
= \sum_{i j} (\frac {2 t^{2}_{ij}}{U_{eff}-V_{eff}} - \beta^{4} J^{H}_{p}) \vec {S_{i}} \cdot \vec {S_{j}},  
\end{eqnarray}
where $U_{eff}-V_{eff} = \alpha^{4} (U_{d} - V_{dd}) + \frac{\beta^{4} J^{H}_{p}}{2}$.
In the case of a nearly 90$^{\circ}$ metal-oxygen-metal bond (Fig.1), there is an 
additional ferromagnetic contribution 
to the total exchange interaction. The origin of this term is Hund's rule exchange interaction on the oxygen atom. 
As we show below, the value of $J_{ij}^{Coulomb}$ is not negligible: it can fully compensate the kinetic contribution, 
so that the total exchange interaction becomes ferromagnetic. 

If $\beta$=0, then the Hamiltonian of Eq.(20) is the simple Hubbard model.
The Coulomb parameters of the general Hamiltonian of Eq.(20) can be calculated 
through constrained LDA calculations or direct
integration over the wave functions. In this paper we use the constrained LDA calculation 
approach, \cite{Andersen} which has given reasonable 
results for a number of compounds.
We apply the analysis of this section to the investigation of exchange interactions in LiCu$_{2}$O$_{2}$.

\section{Results} 
\subsection{LiCu$_{2}$O$_{2}$: LDA CALCULATION}
LiCu$_{2}$O$_{2}$ (Fig.2) is a good example to demonstrate the competition between 
Coulomb  $J^{Coulomb}_{ij}$ and kinetic $J^{kin}_{ij}$
exchange interactions. Indeed, the angle of Cu-O-Cu bond in this compound is equal to 94$^{\circ}$. 
This value of the
angle corresponds to the middle of the ferromagnetic range,\cite{Graaf} and the nearest-neighbor hopping is considerably suppressed (see Eq.(9)).
\begin{figure}[!h]
\includegraphics[width=0.33\textwidth]{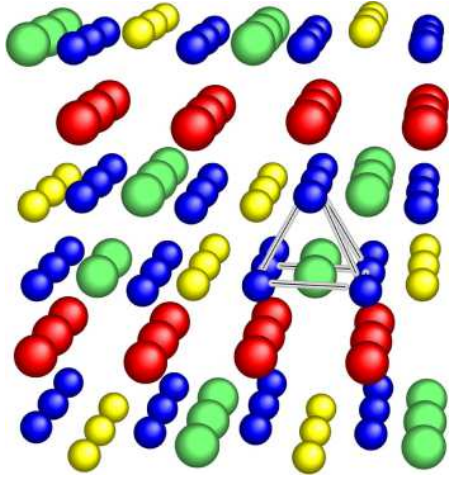}
\caption{(Color online) Crystal structure of LiCu$_{2}$O$_{2}$. Green, red, blue and yellow spheres are Cu$^{2+}$, Cu$^{+}$, O and Li ions,
respectively.}
\label{bandslda}
\end{figure}

D. A. Zatsepin {\it et al.} \cite{Zatsepin} have performed the first {\it ab initio} calculations of the electronic structure of 
LiCu$_{2}$O$_{2}$ in terms of LSDA and LSDA+U approximations. They have concluded that, in the ground 
state, LiCu$_{2}$O$_{2}$ is an insulator with ferromagnetic ordering. 

A. A. Gippius {\it et al.} \cite{GippiusMorozova} have performed magnetic resonance measurements 
of LiCu$_{2}$O$_{2}$ in the paramagnetic and magnetically  ordered states. They have also 
performed full potential LDA calculations. The band structure near Fermi level has 
been fitted by an extended tight-binding model. Using the resulting hopping parameters, the authors
have estimated the values of the exchange interactions $J_{ij}$ within a single band Hubbard model. 

Experimental investigations of the magnetic ordering of LiCu$_{2}$O$_{2}$ by neutron scattering 
have been performed by T. Masuda {\it et al}.
\cite{Masuda,Masuda2} The authors have proposed different sets of exchange constants obtained 
by fitting the calculated spin wave dispersion relation to the experimental curves. None of the
sets of parameters is in good agreement with the results of 
Refs. \onlinecite{GippiusMorozova}. One can conclude that, at present, there is no consistent 
theoretical and experimental description of the
magnetic properties of LiCu$_{2}$O$_{2}$.  

In the present paper, the electronic structure calculation of LiCu$_{2}$O$_{2}$ was performed 
using the Tight Binding Linear-Muffin-Tin-Orbital Atomic Sphere Approximation 
(TB-LMTO-ASA) method in terms of the conventional local-density approximation \cite{AndersenLDA}
and crystal structure data from Ref. \onlinecite{Berger}. 
The band structure of LiCu$_{2}$O$_{2}$ obtained from LDA calculations is 
presented in Fig.\ref{bandslda}. There are four bands near the Fermi level which are well separated from others.
\begin{figure}[!h]
\includegraphics[width=0.33\textwidth,angle=-90]{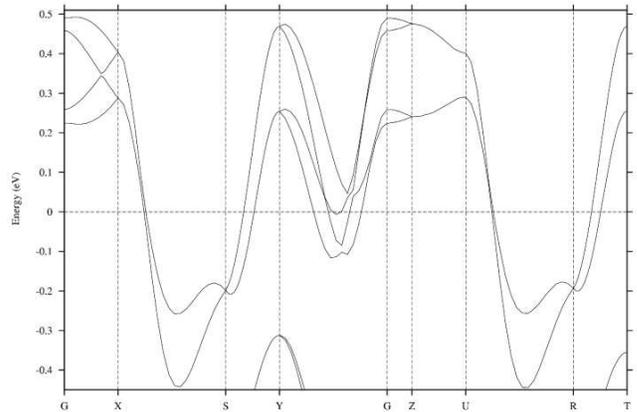}
\caption{Band structure of LiCu$_{2}$O$_{2}$ near the Fermi level (0 eV).}
\label{bandslda}
\end{figure}
These bands are in good agreement with those presented in Ref. \onlinecite{GippiusMorozova}.

\begin{figure}[!h]
\includegraphics[width=0.4\textwidth,angle=-90]{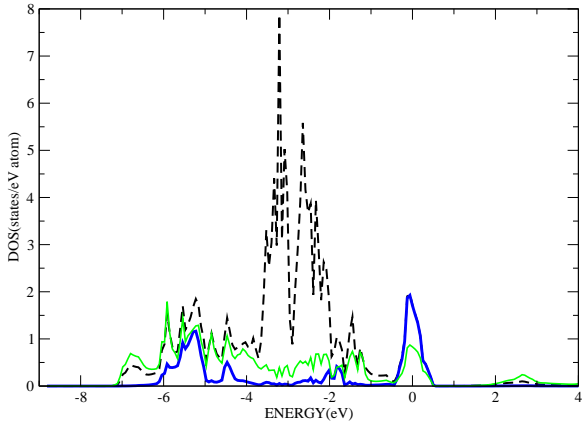}
\caption{(Color online) Partial density of states obtained from LDA calculations. The blue solid and the dashed lines are the density of copper 3d states 
of $x^{2}-y^{2}$ symmetry and the total density of 3d states, respectively. The green solid line is the
density of oxygen 2p states.}
\label{dos3dlda}
\end{figure}

The partial density of states of LiCu$_{2}$O$_{2}$ obtained from LDA calculations is presented in Fig.\ref{dos3dlda}. 
Copper 3d states of $x^{2}-y^{2}$ symmetry are 
strongly 
hybridized with oxygen 2p states. 
Therefore it is more natural to use the Wannier function basis rather than atomic orbitals
to describe the hybridization processes in LiCu$_{2}$O$_{2}$. 
We have used the projection procedure \cite{proj} which is more accurate than the 
fitting procedure used in Ref. \onlinecite{GippiusMorozova} because the Wannier states in the former method 
are constructed from all electron DFT orbitals.
\begin{figure}[b]
\includegraphics[width=0.47\textwidth]{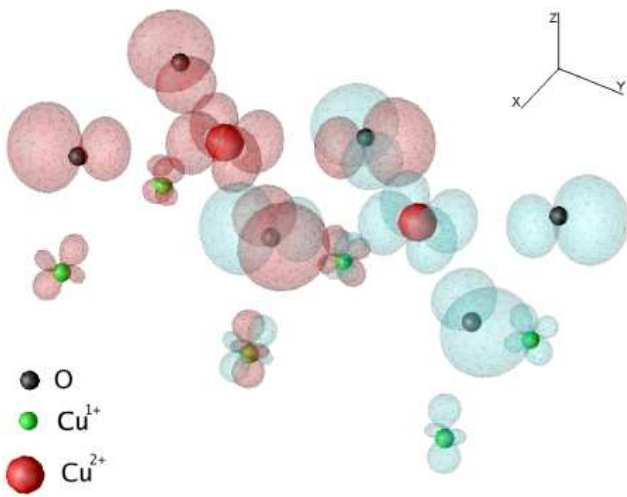}
\caption{(Color online) Wannier orbitals centered on neighboring copper atoms along y axis.}
\label{Wannier}
\end{figure}
The resulting Wannier orbitals of LiCu$_{2}$O$_{2}$, each centered at one Cu site, are shown in Fig.5.  
One can see that the Wannier orbitals strongly overlap 
at oxygen and Cu$^{+}$ atoms. 

As we mentioned above, one of the possible microscopic mechanisms 
of exchange interaction is Hubbard-like AF superexchange. This interaction comes from hopping processes. 
We have calculated the hopping 
integrals between orbitals of x$^{2}$-y$^{2}$ symmetry in the Wannier function basis (Table I).   
\begin{table}[!h]
\centering
\caption [Bset]{Calculated values of hopping parameters $t_{ij}$ between $x^{2}-y^{2}$ orbitals of copper atoms for the one-orbital model and
estimated exchange interaction parameters using Eq.(22) (in meV).}
\label {basisset}
\begin {tabular}{lccccccc}
    \hline
                                            & $y$           &  $2y$       & $x$        & $\tilde x$       & $xy$      & $\tilde {xy}$  & $xyz$\\
  \hline
  $|t_{ij}|$                                &  54           &   99        &  67          &    5.3          & 33     & 28      &  32 \\  
  $|t_{ij}|$ (Ref.\onlinecite{GippiusMorozova}) &  64       & 109         &  73          &     18          &  25    &  -      &   - \\
  $J^{x^{2}-y^{2}}_{ij}$                  &  -24            &  6.5       &  3.0         &   0             & 0.7    & 0.5    &   0.7  \\    
  \hline
  \hline
\end {tabular}
\end {table}
The corresponding interaction paths are presented in Fig.6. 
One can see that for the largest hopping parameters we have good 
agreement with previous band fitting results.\cite{GippiusMorozova} 
However, there are also interaction paths which were not considered before.  

\begin{figure}[b]
\includegraphics[width=0.4\textwidth]{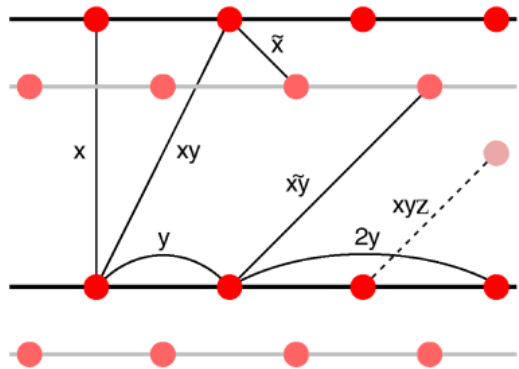}
\caption{(Color online) The schematic representation of interaction paths between copper atoms in LiCu$_{2}$O$_{2}$.}
\end{figure}

For  nearest neighbors along the $y$ axis, there is another contribution to the total exchange interaction, namely
a FM ``direct Coulomb exchange'' between Wannier functions. The simplest physical representation 
of this situation is presented in Fig.7.
\begin{figure}[h]
\includegraphics[angle=270,width=0.4\textwidth]{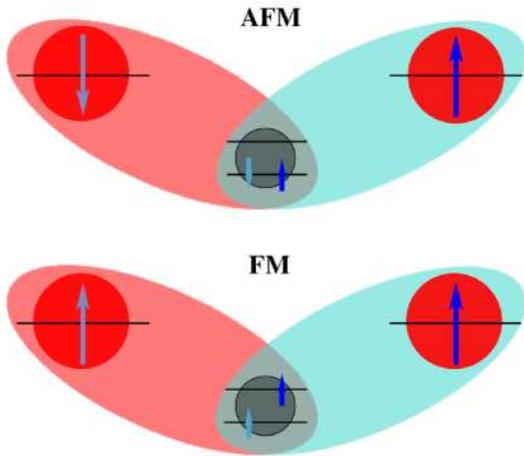}
\caption{(Color online) Schematic representation of AFM and FM configurations of magnetic moments at Wannier orbitals.}
\end{figure}
In the case of the AFM configuration, the magnetic moment of the oxygen atom located between two copper 
atoms vanishes.
By contrast, the magnetic moment of the oxygen atom in the FM case is not zero, and the energy gain is $J^{H}_{p} \times \beta^{4}$, where $\beta$ is the contribution of the oxygen atomic 
orbital to the Wannier orbital (see previous section).
In order to calculate the couplings between magnetic moments, we use the following expression, which comes
directly from the previous section:
\begin{eqnarray}
J_{ij} = \frac {2 t^{2}_{ij}}{\alpha^{4} (U_{d} -  V_{dd}) + \frac{\beta^{4} J^{H}_{p} N_{ox}}{2}} - \beta^{4} J^{H}_{p} N_{ox},
\end{eqnarray}
where $N_{ox}$ is the number of oxygen atoms on which the Wannier orbitals overlap.
For simplicity, we neglect the intersite Coulomb interactions.
The on-site Coulomb and intraatomic exchange interaction parameters of copper atom are determined from 
first-principle constrained LDA calculations: 
$\tilde {U_{d}}$ =10 eV and $\tilde J_{d}^{H}$ =1 eV. 
Therefore the effective Coulomb interaction in Eq.(22) is $U_{d} = \tilde {U_{d}} - \tilde J_{d}^{H}$ = 9 eV. The value 
of the intraatomic exchange interaction of oxygen
atom, $J^{H}_{p}$, was estimated in LSDA+U calculations through the shift of oxygen 2p band centers for spin-up, $C_{\uparrow}$ and 
spin-down, $C_{\downarrow}$: $J^{H}_{p}=(C_{\uparrow}-C_{\downarrow})/M(O)$, where $M(O)$ 
is the oxygen magnetization. 
The obtained value of 1.6 eV is in good agreement with previous estimations. \cite{Mazin}
The value of  $\alpha^{2}$ is related to the magnetization of copper atoms. Our LSDA+U results 
(see Table II) show that $\alpha^{2}$=0.58. 
Since the magnetic moment of the oxygen atom is the result of the magnetization of two copper atoms 
(see Fig.7),  $\beta^{2}$ = M(O)/2 = 0.09.
Using Eq.(22) with the parameters defined above, one can calculate the exchange couplings 
between magnetic moments in LiCu$_{2}$O$_{2}$. 
These results are presented in Table I. One can see that the coupling between nearest neighbors along the $y$ 
axis is strongly ferromagnetic.
These model considerations provide a microscopic explanation of the first-principle LSDA+U results presented in the 
next section.

\subsection{LiCu$_{2}$O$_{2}$: LSDA+U CALCULATION}
The analysis of the previous section shows that one should take into account Coulomb 
on-site correlations and spin polarization of the oxygen atoms.
This has been achieved using LSDA+U. 
The electronic structure of LiCu$_{2}$O$_{2}$ within LSDA+U  is similar to that reported in Ref.
\onlinecite{Zatsepin}. LiCu$_{2}$O$_{2}$ is an insulator with an energy gap of 0.7 eV. The values of 
the calculated magnetic moments (all
values in units of $\mu_{B}$) are 0.58 for Cu$^{2+}$ and 0.18 for O. 

The next step of the investigation is a first-principle calculation of the isotropic exchange integrals for 
the Heisenberg model. In order to calculate the couplings between nearest neighbors along the $y$ axis,
one should use a method which takes into account the change of magnetization of the oxygen atoms. 
The most appropriate one is the method \cite{Moreira} in which the magnetic interaction $J_{ij}$ is estimated 
through the total energy difference between the ferromagnetic and antiferromagnetic first-principle solutions 
(obtained, for instance,  using the LSDA+U approach).
\begin{figure}[h]
\includegraphics[width=0.3\textwidth]{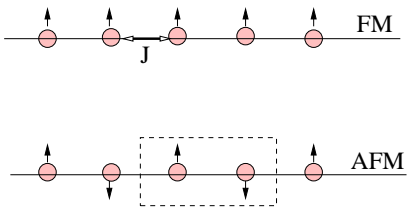}
\caption{(Color online) Ferromagnetic and antiferromagnetic configurations in an infinite chain with identical nearest 
neighbor exchange interaction $J$. The dashed line rectangle denotes the unit cell used for the mapping 
between the Heisenberg model and the first-principle LSDA+U approach.}
\end{figure}
The Heisenberg Hamiltonian describing the interaction between spins in the unit cell (Fig.8) is given by
\begin{equation}
H=2 z J  \vec S_{1} \cdot \vec S_{2},
\end{equation}
where z is number of nearest neighbors.
The corresponding total energies of the ferromagnetic and antiferromagnetic configurations 
of two spins are given by:
\begin{equation}
E_{FM}=  2 z J S^{2}
\end{equation}
and
\begin{equation}
E_{AFM}= - 2 z J S^{2} .
\end{equation}
Therefore the exchange interaction $J$ is expressed in the following form:
\begin{equation}
J = \frac{E_{FM}-E_{AFM}}{4zS^{2}}.
\end{equation}

We have performed calculations for the (1 $\times$ 2 $\times$ 1) supercell in ferromagnetic and
antiferromagnetic configurations. The results are presented in Table II.
\begin{table}
\centering
\caption [Bset]{Results of LSDA+U calculations for the supercell $1 \times 2 \times 1$. M(Cu$^{2+}$) and M(O) 
are magnetic moments of
copper and oxygen atoms, respectively. E$_{tot}$ is the relative total energy (in meV) normalized to 4
(number of copper pairs in unit cell).}
\label {basisset}
\begin {tabular}{cccc}
  \hline
  \hline
                 & M(Cu$^{2+}$)   &  M(O)      & E$_{total}$  \\
  \hline
  FM             &   0.58          &  0.18     &  0            \\
  AFM            &   0.58          &  0        &  38       \\
  \hline
  \hline
\end {tabular}
\end {table}
One can see that in the case of the ($1 \times 2 \times 1$) supercell, the magnetization on oxygen atoms for
the antiferromagnetic configuration is zero, whereas the compensation of magnetization on oxygen atoms in 
the ferromagnetic configuration does not take place. 
Therefore first-principle LSDA+U calculations support the model considerations presented in
the previous section. The ferromagnetic
configuration has a lower energy than the antiferromagnetic one. 
Using Eq.(26) with z=2 and S=$\frac{1}{2}$, we get: $J^{y}_{ij}$ = -19.1 meV.

\begin{table}[!h]
\centering
\caption [Bset]{Values of exchange interactions J$_{ij}$ between magnetic moments of LiCu$_{2}$O$_{2}$ compound (in meV).}
\label {basisset}
\begin {tabular}{cccccccc}
  \hline
  \hline
                       & $y$           &  $2y$       & $x$    & $\tilde x$           & $xy$     & $\tilde {xy}$   & $xyz$ \\
  \hline
  J$_{ij}$  (this work)            &  -19.1   &    9.8         &      3.8          &     0          &  1.0        &   1.0 &  0.4 \\
  J$_{ij}$ (Ref. \onlinecite{GippiusMorozova})    &   -4          &  7.2         &   2.8        &  0.2     &  0.4      & -    & - \\
  J$_{ij}$ (Ref. \onlinecite{Masuda2}, Model 1)    &   -5.95          &  3.7         &   0.9        &  3.2     &  -      & -     & - \\
  J$_{ij}$ (Ref. \onlinecite{Masuda2}, Model 3)    &   -7.0          &  3.75         &   3.4        &  0     &  -       & -  & -\\
  \hline
  \hline
\end {tabular}
\end {table}

In order to calculate the other magnetic interactions, 
we have implemented the Green's function method. \cite{Liechtenstein}
According to this method, we determine the exchange interaction
parameter between copper atoms via the second variation of the total energy with respect to small deviations of the magnetic moments
from the collinear magnetic configuration.
The exchange interaction parameters J$_{ij}$  for the Heisenberg model (Eq.(3)) with S=$\frac{1}{2}$
can be written in the following form: \cite{Liechtenstein,Mazurenko}
\begin{eqnarray}
J_{ij} = \frac{1}{\pi} \int_{-\infty}^{E_{F}} d\epsilon \, {\rm Im} \sum_{\substack {m, m' \\ m'', m'''}}
(\Delta^{mm'}_{i} \,
G_{ij \, \downarrow}^{m'm''} \, \Delta^{m'' m'''}_{j} \, G_{ji \, \uparrow}^{m''' m}) \nonumber
\end{eqnarray}
where $m$ ($m'$, $m''$, $m'''$) is the magnetic quantum number, the on-site potential $\Delta^{mm'}_{i}=H^{m m'}_{ii \, \uparrow} - H^{m m'}_{ii \, \downarrow}$
and  the Green's function is calculated in the following way
\begin{eqnarray}
G^{mm'}_{ij \sigma}(\epsilon) \, = \, \sum_{k,\, n} \frac{c^{mn}_{i \sigma} \, (k) \, c^{m'n \, *}_{j \sigma} \,
(k)}{\epsilon-E^{n}_{\sigma}}.
\end{eqnarray}
Here $c^{mn}_{i\sigma}$ is a component of the {\it n}th eigenstate, and E$_{\sigma}^{n}$ is the corresponding eigenvalue.

Our results are summarized in Table III. One can see that LSDA+U results are in good 
agreement with those obtained in our model analysis (see previous section) and disagree with previous theoretical estimates. \cite{GippiusMorozova} 
This agreement between the model analysis and 
LSDA+U results is very encouraging, but clearly the ultimate test is to compare them with
experiments. 

In that respect as well, the present results are a clear improvement with respect to previous
estimates. Indeed, in contrast to results of paper Ref. \onlinecite{GippiusMorozova}, 
the ratio between the strongest couplings $J^{2y}_{ij} / J^{y}_{ij}$ = -0.5 is in good 
agreement with the results of the neutron scattering 
experiments of Ref. \onlinecite{Masuda2}. This ratio is very important since it controls 
the pitch vector $q$ of the helimagnetic state of
LiCu$_{2}$O$_{2}$. The agreement is not perfect however: Our estimates are about 
twice larger than the integrals deduced from experiments. 
Interestingly enough, it is possible, using the simple microscopic model Eq.(22), 
to identify the source of discrepancy
between LSDA+U and experimental results. Indeed, 
the second term of Eq.(22) is very sensitive to the choice of $\beta^{2}$.
For example, $\beta^{2}$=0.08 leads to $J^{y}_{ij}$ = -18.5 meV, which is in 
excellent agreement with LSDA+U results.
For $\alpha^{2}$=0.65 and $\beta^{2}$=0.06,  we obtain the following set of model 
exchange interactions: $J^{y}_{ij}$=-10 meV, $J^{2y}_{ij}$= 5 meV, 
$J^{x}_{ij}$=2.3 meV, $J^{xy}_{ij}$=0.6 meV, $J^{\tilde xy}_{ij}$=0.4 meV and $J^{xyz}_{ij}$=0.5 meV. 
These model magnetic couplings are in good agreement with the experimental results.

This proves the sensitivity of the results to the precise form of the Wannier
functions. Now, it is well known that several sets of localized functions can be used to 
described a given band, \cite{Lechermann} and the question of which Wannier functions should be used
in the case of the determination of magnetic exchange, a point already
raised by Anderson a long time ago, \cite{Anderson} has not been settled yet. The present
results call for further investigation of that issue. 

Another possible way however to improve the agreement between theory and experiment 
could be the following. 
From the experimental point of view, LiCu$_{2}$O$_{2}$ has spiral magnetic order in the ground state.
Our study was performed for collinear magnetic configurations. Therefore, it would be more natural 
to calculate the exchange couplings using the magnetic structure observed  experimentally. This goes
beyond the scope of the present paper however.

\section{Discussion}
In conclusion, we have presented  an analysis of the Hubbard model in the case of nearly 90$^{\circ}$ 
metal-oxygen-metal bonds dealing explicitly with Wannier orbitals. 
This has allowed us to derive an explicit expression of exchange integrals entirely in
terms of parameters that can obtained from constrained LDA and LSDA+U. This
expression can serve to interpret LSDA+U {\it ab initio} estimates of the exchange
integrals, and to establish their reliability.
The analysis has been applied to the investigation of the magnetic couplings of LiCu$_{2}$O$_{2}$,
allowing us to reach qualitative agreement with experiments, and to gain insight into the
nature of exchange in that system.
Because of the formation of a strongly hybridized, and energetically isolated combination 
of $3d_{x^{2}-y^{2}}$ and 2p orbitals, a large moment is transferred to the O ions, 
and the magnetization of oxygen atoms
has been proven to be the main source of ferromagnetism in LiCu$_{2}$O$_{2}$.

We would like to thank I.V. Solovyev, A.I. Lichtenstein, A.O. Shorikov, J. Dorier and A. Gell\'e 
for helpful discussions.
The hospitality of the Institute of Theoretical Physics of EPFL is gratefully acknowledged.
This work is supported by INTAS Young Scientist Fellowship Program Ref. Nr. 04-83-3230, 
Netherlands Organization for Scientific Research 
through NWO 047.016.005, 
Russian Foundation for Basic Research grant RFFI 04-02-16096, RFFI 06-02-81017 and the grant program of President of Russian Federation
Nr. MK-1573.2005.2.
The calculations were performed on the computer cluster of ``University Center of Parallel Computing'' of USTU-UPI. 
We also acknowledge the financial support of the Swiss National Fund and of MaNEP.

\end{document}